\documentclass{aip-cp}

\usepackage[numbers,sort&compress]{natbib}

\usepackage{rotating}
\usepackage{graphicx}
\usepackage{sidecap}
\sidecaptionvpos{figure}{t}

\usepackage{wrapfig}

\newcommand{\un}[1]{\>\mathrm{#1}}

\begin{document}

\title{Laser Ablation of Gold into Water: near Critical Point Phenomena and Hydrodynamic Instability}

\author[aff1,aff2]{Nail Inogamov \corref{cor1}}
\author[aff2,aff1]{Vasily Zhakhovsky}
\author[aff1]{Viktor Khokhlov}

\affil[aff1]{Landau Institute for Theoretical Physics, Chernogolovka 142432, Russia}
\affil[aff2]{Dukhov Research Institute of Automatics, Moscow 127055, Russia}
\corresp[cor1]{Corresponding author: nailinogamov@gmail.com}

\maketitle

\begin{abstract}
Laser ablation of gold irradiated through the transparent water is studied.
 We follow dynamics of gold expansion into the water along very long (up to 200 ns) time interval. This is significant because namely at these late times pressure at a contact boundary between gold (Au) and water decreases down to the saturation pressure of gold.
 Thus the saturation pressure begins to influence dynamics near the contact.
 The inertia of displaced water decelerates the contact.
 In the reference frame connected with the contact, the deceleration is equivalent to
   the free fall acceleration in a gravity field.
 Such conditions are favorable for the development of Rayleigh-Taylor instability (RTI) because heavy fluid (gold) is placed above the light one (water) in a gravity field.
 We extract the increment of RTI from 2T-HD 1D runs.
 Surface tension and especially viscosity significantly dump the RTI gain during deceleration. Atomistic simulation with Molecular Dynamics method combined with Monte-Carlo method (MD-MC) for large electron heat conduction in gold is performed to gain a clear insight into the underlying mechanisms. MD-MC runs show that significant amplification of surface perturbations takes place.
 These perturbations start just from thermal fluctuations and the noise produced by
   bombardment of the atmosphere by fragments of foam.
 The perturbations achieve amplification enough to separate the droplets from the RTI jets of gold. Thus the gold droplets fall into the water.
\end{abstract}

\section{1. INTRODUCTION}

We consider the problem of dynamics of gold illuminated through water by ultrashort laser pulse.
 This problem is interesting itself due to its complexity
  and it is important for nanotechnological applications connected with a clean way (without chemistry)
    of nanoparticles production and also for creation of functional surfaces
      (e.g. for enhancing of surface Raman scattering) which differ from the functional surfaces
        produced by illumination through vacuum or gas.
 We begin with short presentation of the two-temperature phenomena
   inevitable when the ultrashort laser pulse is used.
 We present results of two-temperature (2T) one-dimensional hydrodynamic (2T-HD) simulations
  covering very long (up to 0.2$\mu$sec) time interval.

 This is significant because namely at these late times pressure at a contact boundary
   between gold (Au) and glass decreases down to saturation pressure of gold.
 And the saturation pressure begins to influence dynamics near the contact.

 Inertia of water is the next main actor.
 It decelerates the contact.
 In the reference frame connected with the contact the deceleration is equivalent to
   the free fall acceleration in a gravity field.
 This follows from the Einstein's principle of the gravity/inertia equivalence.
 This is exact the conditions favorable for development of Rayleigh-Taylor instability (RTI)
  because heavy fluid (gold) is placed above the light one (water) in a gravity field.

 We extract the increment of RTI from 2T-HD 1D runs.
 Surface tension and especially viscosity significantly dump the RTI gain during deceleration.

 We use large scale molecular dynamics simulations combined with Monte-Carlo approach (MD-MC) for electron heat conduction to do the situation clear.
 MD-MC runs show that significant amplification of surface perturbations takes place.
 These perturbations start just from thermal fluctuations and the noise produced by
  bombardment of the atmosphere by fragments of foam.

 The perturbations achieve amplification enough to separate the droplets from the RTI jets of gold. Thus the droplets fall into water.

 There is a quasi-hydrostatic equilibrium near the contact in gold.
 Therefore we use the word ``atmosphere''.

 Laser action should be strong to produce nanoparticles.
 It is significantly higher than the nucleation threshold $F_{abl}$ for gold
   thermo-mechanically ablated into vacuum.
 Absorbed energy $F_{abs}$ is of the order of or higher than the evaporation (ev) threshold $F_{abs}|_{ev}$
   above which the spallation plate cannot form during expansion of gold to vacuum.
 In this case very wide foamy zone is created.
 Expansion of foam doesn't ``know'' about water.
 Foam expands freely.
 Thus its expansion velocities begin overcome velocity of a contact decelerated by water.
 This causes accretion of membranes of foam onto atmosphere created thanks to deceleration.
 The MD-MC simulations beautifully illustrate this flow
   with shock in water, atmosphere ``sitting on water'', vast foam, RTI of the contact,
    and accretion of foam onto atmosphere.


 Laser ablation of metal in contact with liquid differs much from ablation into vacuum \cite{2004-air-vacuum_fs_Ni-Japan,Povarnitsyn:2013,PovarnitsynItina:2014,LZ:2017a,LZ:2017b}.
 In spite of importance of this type of laser-matter interaction (e.g., for nanoparticles production), the involved processes are still poorly understood.
 But a lot of experimental works were already performed \cite{Shafeev:2009,Shafeev:2009-NP-tail,Barcikowski:2009,Shafeev:2012,Shafeev:2014}
  -- see also recent extensive reviews in \cite{Xiao:2017,Goekce+Barcikowski:2017}.


 We show that to produce nanoparticles the laser absorbed energy $F_{abs}$ should exceed
  the ablation threshold $F_{abs}|_{abl}$ into vacuum by a few times;
    $F_{abs}|_{abl}\approx 100 \un{mJ/cm^2}$ for gold \cite{Inogamov:2008jetp,Norman:2012}.
   Here the ablation with  $F_{abs} = 400\un{mJ/cm^2}$ is studied.
 As a result of large energy absorbed in gold
   the temperatures in the heat-affected zone (HAZ) increase above the critical temperature.
 The flow of the substances, including propagation of a strong shock in liquid
   and a rarefaction wave inside the metal target, is analyzed.
 We demonstrate that the contact between metal and liquid, both being in their supercritical states, is unstable, which leads to the Rayleigh-Taylor instability.
 Dynamics of the instability is important for separation of melt droplets,
   which are frozen to solid nanoparticles later.


 There are a chain in time of interrelated physical phenomena.
 Processes begin with absorption of ultrashort pulse (durations $\sim 0.1-1$ps) and two-temperature (2T) stage $T_e \gg T_i$
     covering duration of a pulse and finishing when electron $T_e$ and ion $T_i$ temperatures equalize.
 Our approach for this stage was described earlier
  \cite{Inogamov:2015a,Ashitkov:2016Elbrus}
   therefore we omit its description in this paper.
 The heat-affected zone (HAZ) is formed during the 2T stage.
 Its thickness for gold is $d_T\approx 150$nm.
 The HAZ is formed inside gold independently of presence or absence of water.


 The chain of processes proceeds with acoustic decay of HAZ, see Section 2.
 In the case of high $F_{abs}$ considered here,
   vast region occupied by foam is formed as a result of decay of HAZ.
 Foam is a mixture of liquid and vapor phases with low volume content of liquid in a unit volume of mixture.
 Foam develops during expansion of zone of nucleations.
 Acoustic impedance $z_{wt}$ of water is small relative to that for gold $z_{Au}/z_{wt}\approx 40.$
 Therefore presence of water is an insignificant factor at the stage of decay of liquid condensed Au
  into two-phase mixture; this stage is studied in Section 2;
    the stage lasts during an acoustic time scale $t_s=d_T/c_s,$ where $c_s$ is speed of sound.
 But later in time the main part of foam accretes onto atmosphere appearing namely due to presence of water. 

   \section{2. DECAY OF METAL INTO VACUUM OR LIQUID}

\begin{SCfigure}[\sidecaptionrelwidth][!ht]
  \centering
\includegraphics[width=0.6\textwidth]{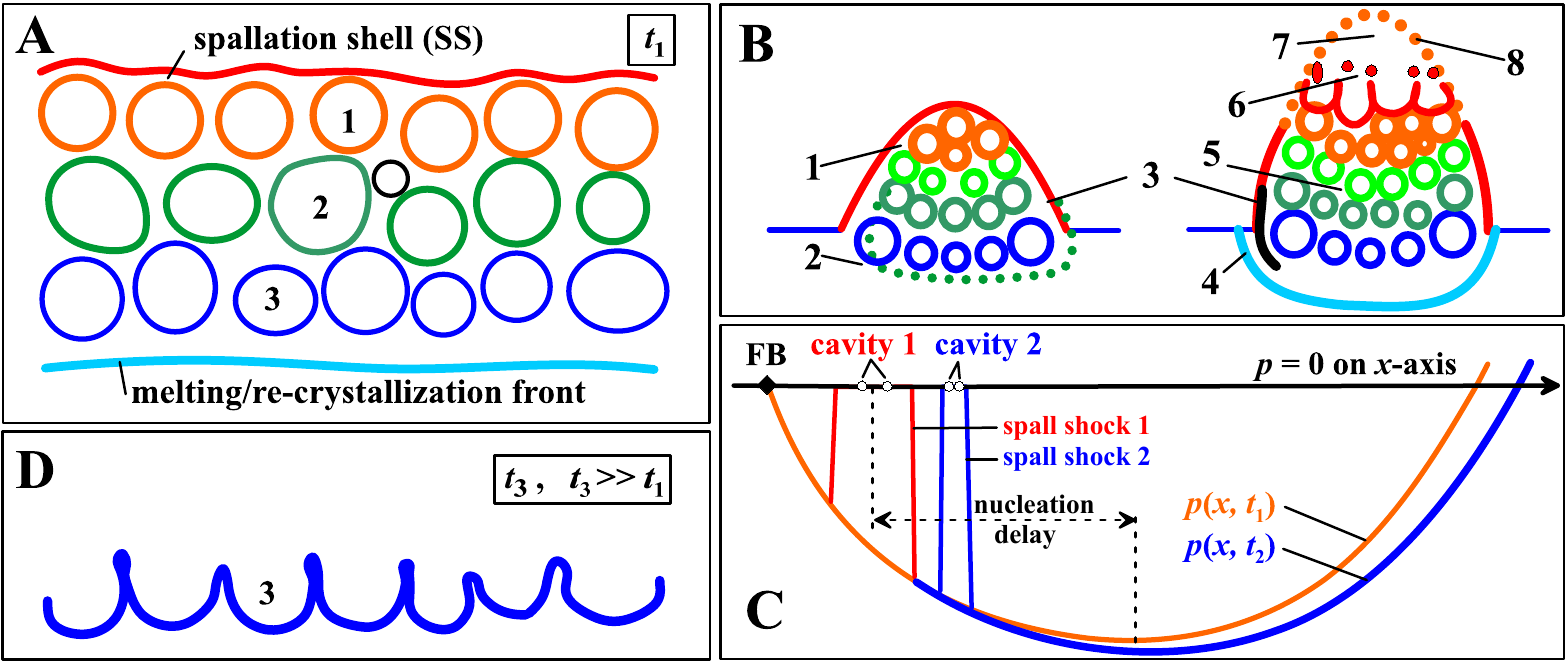} 
\hspace{0.01\textwidth}
  \caption{   \label{fig:1} {\bf A.} Melting front, foam, and SS.
  {\bf B.}~Systems of finite lateral sizes $R_L$ are shown;
   the frames A and D correspond to the radius $R_L=\infty.$
  The systems in frame B consist from cupola SS ``1'' which is closed
    if $F_{abs}|_{abl} < F_{abs} < F_{abs}|_{ev}$
   and open in central region ``7, 8'' if $F_{abs}|_{ev} < F_{abs}.$
  {\bf C.}~Nucleation for fluences greatly above the ablation threshold; supersonic propagation of the nucleation front.  This is the main case considered below.
   This is a start stage for processes producing the reach multilayered foam shown in frames A and B, which then finish with the structured surface in frame D.
   This regime is favorable to production of nanoparticles in vacuum or liquid. } 
\end{SCfigure}


 There are two type of experiments with different lateral size of an irradiated spot (radius $R_L)$   at a surface.
 People use diffraction limited tight focusing when $R_L\sim \lambda$ in the first type of experiments
    \cite{Kuchmizhak:2016a,Inogamov:2016APAnanoBump,Danilov:2016,Inogamov:2016NRLnanoBump,Danilov:2016-nanoBump,Wang:2017-PhRevAppl};
      for optical lasers $\lambda\sim 1$$\mu$m.
 While in the other type the large spots $(R_L$ is many microns) are necessary.
 For massive nanoparticle production and for creation of functional surfaces the large spots are used.
 Functional surfaces are covered by random nanostructures.

\begin{SCfigure}[\sidecaptionrelwidth][!ht]
  \centering
\includegraphics[width=0.65\textwidth]{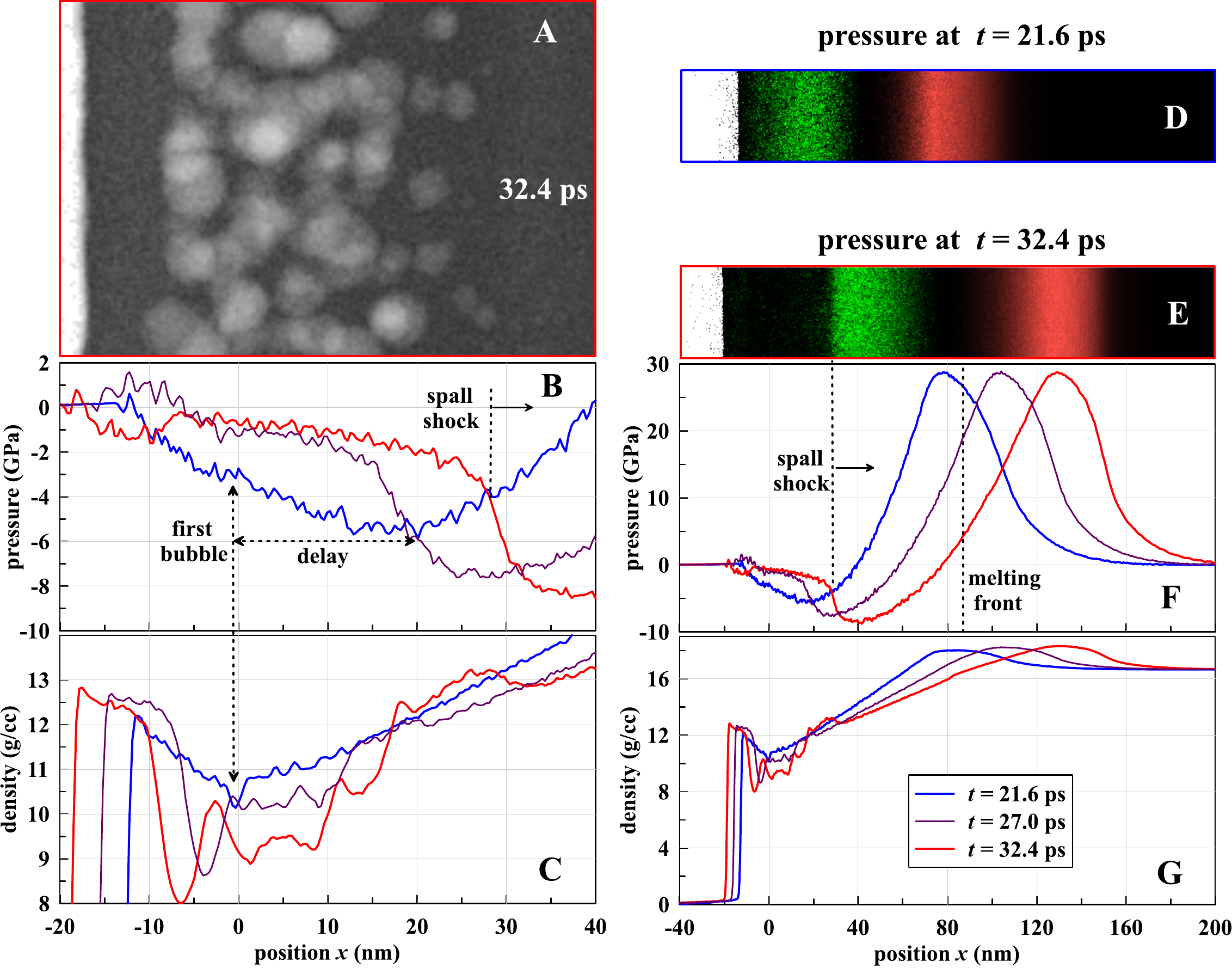} 
\hspace{0.01\textwidth}
  \caption{Simulation of ultrashort laser action onto tantalum sample with dimensions $400.1\times 40.2 \times 20.1 \un{nm^3}$, number of atoms $17.9\times 10^6$,
  HAZ 40 nm (Gaussian distribution), $F_{abs}=275 \un{mJ/cm^2}.$
  {\bf A.} The map of density integrated along the ray perpendicular to the plane of picture.
  We see the ensemble of the 3D cavities; neither of them forms through hole
   (through thickness of simulation box in direction perpendicular to the plane of picture),
    therefore this is the 3D flow.
  {\bf F.} The pressure profiles in whole.
  They are repeated in frames {\bf D} and {\bf E} as the pressure maps -
    the colors of those bounding frames corresponds to the colors of the curves presenting the profiles.
  {\bf D, E.} We see the compression (red color) and rarefaction waves (green) together with unloaded region
   with small pressures (black color) in Figure~E.
  The right edge of the left black region in Figure~{\bf E} relates to the spall shock.
  {\bf B, F} and {\bf C, G} are pressure and density profiles, resp.}
\end{SCfigure}


 For large spots the surface tension   defines spatial scales of frozen random structures placed at the bottom of a crater  (see Figure 1D)  and observed in experiments
   after a single or few shots action;
      many peoples saw them
 \cite{Vorobyev:2006,Fang:2017,Inogamov:2015b,Gurevich:2011,Ashitkov:2015,InogamovCCP:2014}
        and have advanced useful applications of these structured surfaces          for changing optical 
         \cite{Vorobyev:2006,Fang:2017},
             mechanical (strength, tribology, wettability) \cite{Vorobyev:2006},
                or catalytic properties of a surface.
 These capillary scales are much smaller than diameter of a large crater
    \cite{Gurevich:2011,InogamovCCP:2014}.
 The surface structures form during expansion and breaking of a liquid-vapor foam,
  see Figures 1C (appearance of nuclei), 1A (development and expansion of nuclei),
 1D (final state formed after breaking and freezing), and  \cite{InogamovCCP:2014}.


 Significantly above ablation threshold a deep foam layer appears.
 This is illustrated in Figure 1A by the bubble layers ``1-2-3'' in the foam.
 Nucleation of the layer ``1'' proceeds first. How the first nuclei appears
   and how they influence rarefaction is shown in Figure 1C and in Figure 2 below.
 The next (second in Figure 1C - cavity 2) layer of cavities appears {\it ahead} the spall shock 1,
  that is the cavity 2 nucleates in the region of strong stretching -
   a spallation shock eliminates the stretching, increases local density, and thus prevents nucleation,
    see also additional discussion below.
 The fact that the ``cavity 2'' overruns the ``spall shock 1'' in Figure 1C means that the nucleation front
   runs to the right side faster than speed of sound - this is the phenomenon of supersonic nucleation.


 There is a spallation shell above the bubble layers    in the case $F_{abs}|_{abl} < F_{abs} < F_{abs}|_{ev}$
    \cite{Inogamov:2008jetp,Agranat:2007}.
 The cupola in Figure 1B is filled with expanding multi-layer foam.
 Above the ``evaporation'' (ev) threshold $F_{abs}|_{ev}$
   the integral (this means that the shell begins at the unperturbed target surface and continues up to the tip) shell SS
     covering the foam in Figure 1A disappears \cite{Inogamov:2008jetp,Agranat:2007}.
 The left cupola in Figure 1B has integral SS while the right one has a SS with the hole 7, 8 around its tip.


 Bubbles develop in molten metal,    therefore at the early stages the melting front in Figure 1A is below the foam.
 Later in time the shell and membranes in the upper bubble layers break off and fly away.
 The remnants of the deep layer ``3'' in Figure 1D remain and solidify at the bottom of a crater ``2'' shown in Figure 1B.
 These remnants form the random relief in Figure  1D.
 This solid relief is observed in experiments in vacuum
    \cite{Vorobyev:2006,Fang:2017,Inogamov:2015b,Gurevich:2011,Ashitkov:2015,InogamovCCP:2014}
       and in the case of ablation in liquid, see Figure 8 in \cite{Shafeev:2009-NP-tail}.


 In Figure 1B
  the bottom of the crater remaining at the surface is marked by ``2'', this bottom is shown in frame D.
  The ring structure around the crater is ``3'' in Figure 1B.
  Melting/re-solidification front is ``4''.
  The bubble interior is ``5''.
  The ``6'' in Figure 1B is a vapor-droplet zone ahead the closed bubbles separated from each other by membranes.
  With rupture of membranes the bubble zone gradually transforms to the vapor-droplet zone.


\begin{figure}
  \centerline{\includegraphics[width=1.\textwidth]{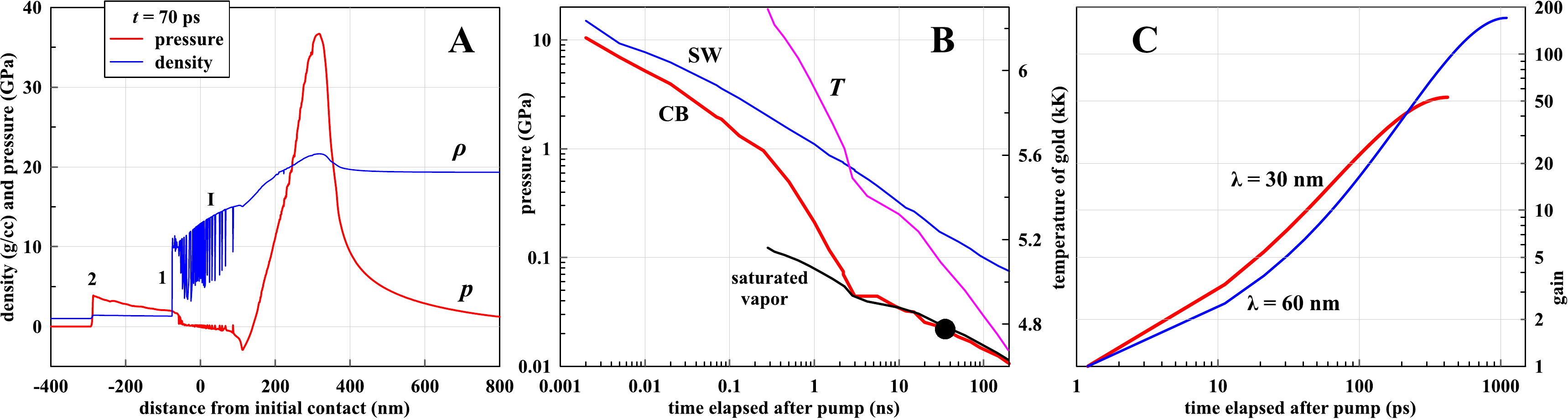}}
  \caption{Simulation of expansion of gold into water; 2T-HD code, $F_{abs}=400 \un{mJ/cm^2},$
 {\bf A.} Profiles of compression wave propagating to the right.
 The wave becomes steeper and overturns forming a shock.
 Cavitation zone is important for gold-water interaction. Pressure in this zone is small.
 The cavitation zone is wide (covering all HAZ) because energy is large.
 Small increase of density and pressure at the left end of cavitation zone corresponds to the atmosphere.
 It is supported by contact pressure.
 {\bf B.} Decay in time of pressures at shock in water (curve ``SW'') and at the contact boundary ``CB''.
 Contact pressure drops down to saturation pressure in gold at temperatures of atmosphere shown by
  the curve ``T''. Further drop decreases contact pressure down to critical pressure of water (black circle).
 {\bf C.} Linear gain of perturbation amplitudes
  calculated from known from simulation dependence of deceleration $g(t)$
   and values for surface tension and viscosity of gold taken from reference books.
 Here we show typical perturbation wavelengths 30 and 60 nm.
  }
\end{figure}


 Ring shape around a crater appears in competition between capillarity and re-solidification,
   see Figure 1B where it is marked by ``3''.
 The ring corresponds to a region near the thermomechanical threshold
   where (around the ring) small open and closed undersurface bubbles locate
     \cite{Ashitkov:2012,Wu:2015}.
 Development of a crater may be imaged as inflation of a large bubble (called cupola above)
  together with inflation of foam filling this bubble,
    and successive ruptures of membranes forming foam
     beginning from the upper layers of bubbles, see Figures 1A, 1D and 1B.


   \section{3. NUCLEATION AND FOAM - FORMATION AND DECAY}


 Capillary velocities $v_\sigma$ are rather low, usually $\sim$10 m/s.
 But surface tension phenomena, e.g. the final crater nanostructures, aren't excluded if laser action is strong
  and expansion velocities are much higher than $v_\sigma;$
    they are not excluded if a film is thick enough to keep a cold interior.
 In bulk targets a strong action (i.e. significantly, at least few times, above ablation threshold)
   drives a cascade of nucleations shown in Figures 1 and 2.
 Figures 1C and 2 illustrates how the layers of bubbles (cavities) in the cascade of layers form.
 There are successive nucleation ``flashes'' as a rarefaction wave ``sitting'' at the back side of the compression wave
    penetrates into a target.
 The scheme of successive ``flashes'' is presented in Figure 1C.
 The compression wave going to the right side is shown in Figure 2F.
 Later in time the compression wave breaks off (wave breaking) forming a shock; this late time evolution isn't shown here.


 Nucleation event produces a spherical shock expanding around a cavity.
 The first cavities are located along a plane layer shown in Figure 1C as ``cavity 1''.
 Spherical waves from the individual cavities interfere
  and form two plane spall waves propagating to the left and to the right side
   relative the layer of cavities.
 The right shock is shown as ``spall shock 1'' in Figure 1C.
 Two small circles ``cavity 1'' mark off the ``banks'' of the rupture
   where in the Lagrangian 1D simulation density drops to zero;
     an empty gap between two neighboring Lagrangian particles appears.
 In MD-MC there is a well in density profile between these banks, see Figure 2B and 2C
   where the first well is marked by the two side arrow.
 The arrow shows the point of the first nucleation where density decreases, while pressure increases.


 The spall shock unload tensile stress increasing pressure from negative value to the value $p=0$
   as it is shown on pressure profile $p(x,t_1)$ in Figure 1C by the two vertical straights around the ``cavity 1'';
    the right straight is underlined by caption ``spall shock 1''.
 The ``spall shock 1'' propagates to the right side with speed
   which slightly surpasses local speed of sound $c_s;$ dispersion of $c_s$ depending on density.
 The excess above $c_s$ is proportional to a nonlinearity degree of a spall shock.
 Nonlinearity is weak therefore the excess is small.

\begin{SCfigure}
  \centering
\includegraphics[width=0.54\textwidth]{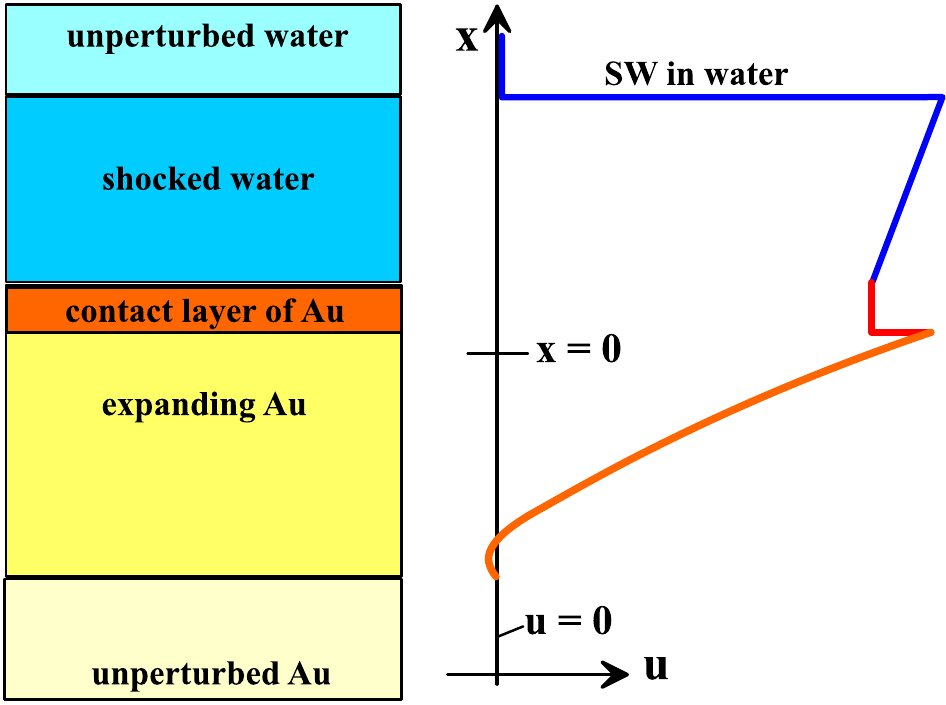}
  \hspace{0.01\textwidth}
  \caption{Layered structure formed after launch of expansion triggered by ultrafast laser impact.  It consists from: (i) the unperturbed water;
  (ii) shocked water layer bounded from above by the shock in water
    and from below by the contact between blue water and gold;
  (iii) the orange layer is the liquid continuous gold engaged in dynamical contact with water;
  (iv) the yellow layer below the contact Au layer is gold inside the expansion (rarefaction) wave,
    see Figure 3A;
  (v) the layer between rarefaction and unperturbed gold
     corresponds to the compression wave running into bulk region of a target, see Figure 3A;
  (vi) the unperturbed gold is waiting for arrival of the pair from compression and rarefaction waves.
  The sequence (i-vi) is is the structure before the nucleation stage and formation of foam.
  After appearance of foam the layer (iv) is changed qualitatively.
  The plot at the right side shows distribution of longitudinal velocity.
  Mainly it is directed to the side of water, only small part below has opposite velocities
    connected with compression wave running into bulk.
  At the velocity profile there are downward: the compressed water - the contact layer - the rarefaction wave.
  Decrease of velocity inside shocked water means that the water shock is decelerating -
   see Figure 3B showing the decrease of pressure ``SW'' at the shock front in water due to its deceleration.
  There is a jump of velocity at the boundary separating the contact layer (called atmosphere in the text) and rarefaction.
  This means that liquid from rarefaction adjusts to the contact layer.
  }
\end{SCfigure}

\begin{SCfigure}[\sidecaptionrelwidth][!ht]
  \centering
  \includegraphics[width=0.54\textwidth]{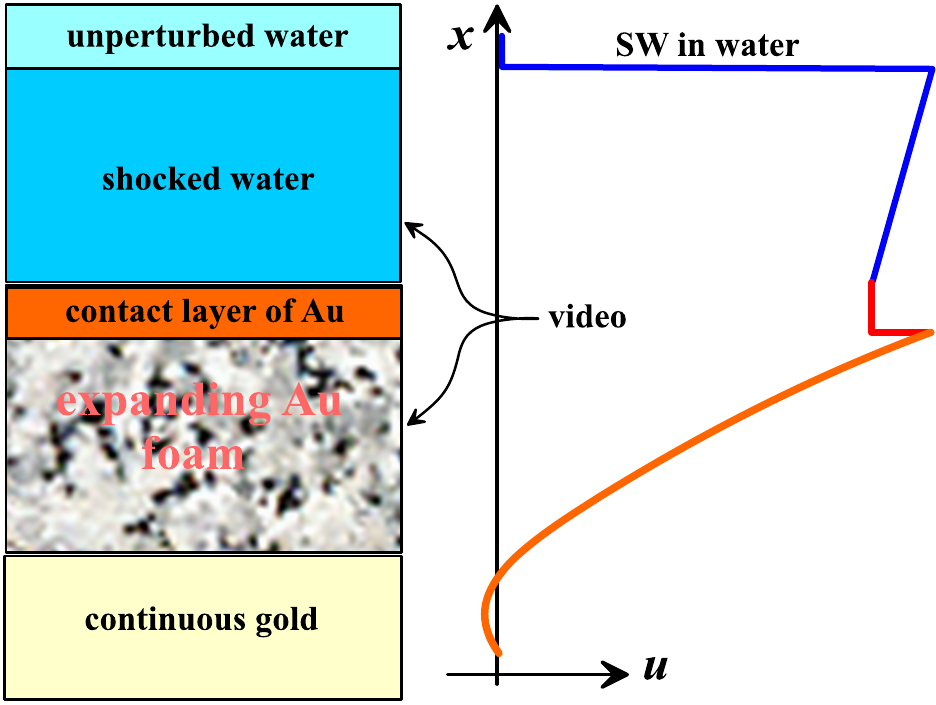} 
  \hspace{0.01\textwidth}
  \caption{This is the structure shown above in Figure 4 but after formation of foam.
  Now the foam accretes (adjusts) to the contact layer.
  This follows from excess of velocity of foam at the boundary between the contact layer and foam shown in velocity profile.
  At the stage with foam, the compression wave is far away from the near contact interaction region,
   it leaves low pressure, motionless gold behind,
     and thus it doesn't influence the near contact dynamics.
  The bracket seizes the most dynamically important region consisting from (i) the near contact water;
   (ii) the contact layer - atmospheric layer from Au in quasi-hydrostatic equilibrium;
   and (iii) foam accreting into the atmosphere.
  Temporal evolution of the layers seized by the bracket are shown in Figure 6.
  }
\end{SCfigure}


 Non-trivial is the regime of multiple nucleations in the successive layers.
 The next nucleation takes place before the spall shock
   propagating from the previous event
     arrives to the point of the next nucleation.
 The ``cavity 2'' in Figure 1C appears to the right side relative to ``spall shock 1'',
   i.e. outside the unloading influence of the ``spall shock 1''.
 The ``cavity 2'' event produces its own ``spall shock 2''.
 Thus the spall shock front during multiple nucleations
  loses its causal relationship with ``old'' nucleation events (as was mentioned in Chapter 2 above).
 In such regime the spall shock runs with phase velocity.
 This velocity surpasses speed of sound but not in connection with $c_s$ dispersion related non-linearity
   as in the case of a single nucleation layer.


 We see that the current position of a spall shock is defined by the last nucleation layer.
 This is true while the nucleation zone develops into bulk matter capturing new and new masses.
 When nucleation ceases then the spall shock returns to its unsupported (by successive nucleations),
   weakly non-linear regime of propagation.


 The well of $p<0$ in the pressure profile shown in Figure 2F has two branches
  to the left and right sides relative to the instant minimum.
 As a rarefaction wave propagates to the right - the minimum and the right branch shift to the right,
   while the left branch remains approximately the same, only minimum becomes deeper,
     thus the left branch is prolonged more far down - up to the current minimum,
      compare the profiles $p(x,t_1)$ and $p(x,t_2>t_1)$ in Figure 1C.
 This is the situation without nucleations.
 During this process of lowering of the minimum, the beginning of the left branch
   is pinned to the point ``FB'' (frontal boundary) in Figure 1C;
    FB means laser irradiated free surface of condensed phase bordering with its vapor.
 At acoustic time scale $t_s=d_T/c_s$ the process of lowering of the minimum terminates;
   here $d_T$ is thickness of a heat affected zone (HAZ); the HAZ is created during a two-temperature stage.
 Then the lowering finishes, the beginning of the left branch comes uncoupled with the free surface
   and begins to move with speed of sound into interior.
 Pressure at the interval connecting the free surface and the moving beginning equals to vapor pressure.


 This was the situation without nucleation.
 At the nucleation threshold the first and last layer of cavities appears in the minimum of the pressure profile
   when the minimum achieves its deepest point.
 We are interested in the case significantly (few times and more) above the nucleation threshold.
 Then the first plane layer of cavities (``cavity 1'' in Figure 1C)
   (i) nucleate at some distance ``nucleation delay'' behind the current position of a minimum in Figure 1C,
     while (ii) the minimum continues to get deeper into $p<0$ region.
 It seems that this is the same minimum of the ``pure'' rarefaction wave
   unaffected (therefore it is said ``pure'') by the ``wave'' of successive nucleation events
     coming from the left side relative to the current minimum.
 At least in Figures 2B and 2F the plum and red minimums look like ``sitting'' at the continuation
  of the blue profile weakly influenced by nucleation.


 In the simulation shown in Figure 2 the cascade of nucleations ceases near the instant $32$ ps
   before the instant $t_*\sim t_s$ when the minimum achieves its deepest depth.
 The right bubbles in Figure 2A are the last ones.
 In our conditions the thermo-fluctuating mechanism of nucleation works.
 Probability to born a nucleus is proportional to $\exp(-W/T),$ $W=(16\pi/3) \sigma^3/p^2,$
   where $W$ is work necessary to born a nucleus, $\sigma(T)$ is surface tension.
 Ending of nucleations is caused by decrease of temperature $T$ with distance from the free surface
   as we approach the melting front shown in Figure 2F.


 Spall shock transfers into its usual (unsupported by nucleations) propagation regime
   after finishing of appearing of new bubbles.
 At late stages the spall shock neutralizes the well of negative pressure because it moves
  slightly faster than the local acoustic characteristics of the rarefaction wave
   because of its non-linear propagation (slightly above the local current speed of sound).


\begin{figure}
  \centerline{\includegraphics[width=0.96\textwidth]{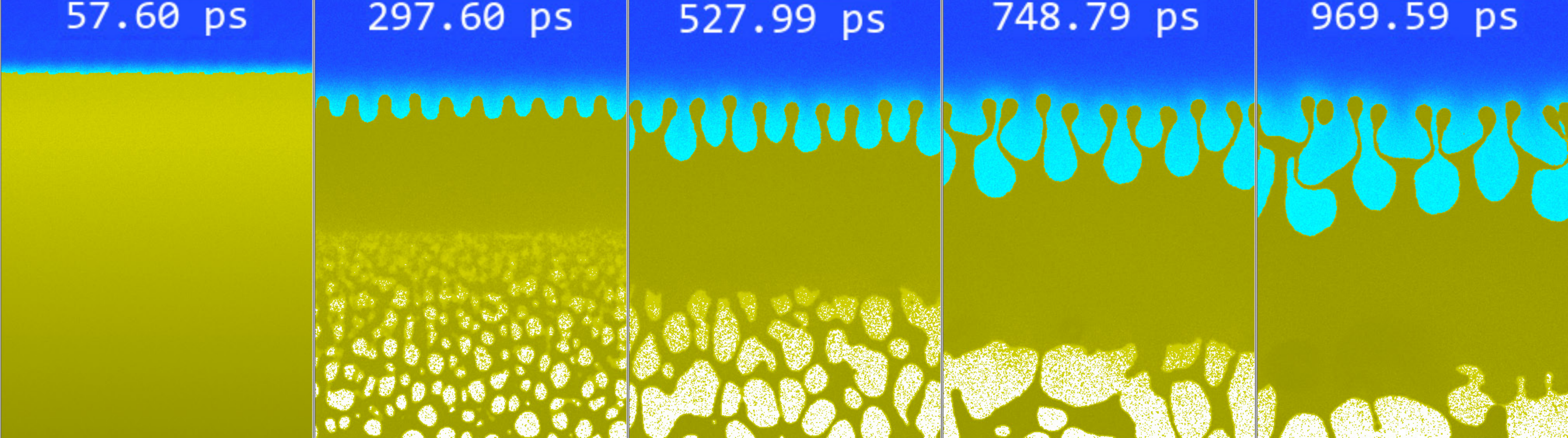}}
  \caption{Evolution of the near contact situation; water is blue, gold is yellow.
  We see formation and development of foam, due to inflation the cells in foam enlarge,
   contact is decelerated by water, a layer of continuous gold appears at the contact,
    we call ``atmosphere'' this layer (because it is hydrostatically supported by water),
     atmosphere decelerates, thus it velocity decreases, while foam ``doesn't know'' about water
      and expands freely - this means with larger current velocity than atmosphere.
  Therefore foam accretes onto the atmosphere.
  Water is heated through atom-atom conduction from hot gold.
  Heated water has less blue color. It is located in growing bubbles near the contact.
  Due to deceleration of dense material (gold) by light one (water) the contact is unstable to
   Rayleigh-Taylor instability (RTI).
  Therefore the well known structure made from bubbles and jets deforms the contact.
 Non-linear development of this structure causes appearance of gold droplets in water.
  }
\end{figure}


 Hydrodynamic velocities of foam expansion
   decrease from external (\#1 in Figure 1A) to internal (\#3 in Figure 1A) successive layers ``1-2-3''
      which make the cascade \cite{Fang:2017,InogamovCCP:2014}.
 Foam develops from the nucleated vapor cavities as they inflate decreasing a volume part of liquid
   in the liquid-vapor mixture.
 Near the practically motionless bottom of a crater, the velocities in the bottom layer \#3 in Figure 1A
    drops down to $v_\sigma.$
 This condition corresponds to the last layer where propagation of a nucleation zone into a target ceases
  (instant $t_1,$ Figure 1A).
 Later in time, capillary remnants of this layer remain on a target, freeze, and form a final surface topology,
  while the spallation shell together with the upper bubble layers
   separate from a target and fly away (instant $t_3,$ Figure 1D) \cite{InogamovCCP:2014}.


 The spallation shell isn't created above the evaporation threshold $F_{ev}$
   introduced in \cite{Inogamov:2008jetp,Agranat:2007};
     see also classification of 3 regimes of expansion between two thresholds
      (ablation and evaporation) in Figure 14 \cite{Inogamov:2008jetp},
       Figure 16 \cite{Wu:2014}, and in \cite{Zhigilei:2009}.
 The spallation shell isn't created because the upper layer is too hot -
   high temperature almost totally suppresses cohesion attraction between atoms.
 In the case $F>F_{ev}$ a plasma or hot vapor ejecta precedes the expanding bubble layers.
 The cases with and without a spallation shell are compared in Figure 1B.


 Evolution of cavitated fluid is shown in Figure 1A and 1D.
 Rarefaction wave runs with speed of sound
  (leaving the trace with successive nucleation layers, see Figure 1C and 2)
     while hydrodynamic expansion of foam proceeds slowly.
 Therefore bubbles in the successive layers ``1-2-3'' in Figure 1A differ moderately in their sizes,
  see also Figure 2A;
   although the layer ``1'' appears before the all successive layers in Figure 1A.
 At $t_1$ metal is molten above the melting/re-crystallization front in Figure 1A.
 The final surface structure solidifies before the instant $t_3$ in Figure 1D.
 This structure is a frozen trace of the last layer of bubbles.
 Foamy matter confined inside the spallation shell is shown in Figure 1B.
 In the case with hot central region $F>F_{ev}$
   the spallation shell ``1'' isn't geometrically closed (this is the right cupola in Figure 1B) -
     there is a hole 7, 8 in the tip region
  \cite{Agranat:2007,Inogamov:2008jetp,Wu:2014}.

  \section{4. HYDRODYNAMIC SIMULATION OF GOLD/WATER DYNAMICS}


 Figure 3 presents results of long (up to $0.2\un{\mu s}$) simulation done by the 2T-HD code.
 Long duration of the run allows to achieve the stage when pressure at the contact drops to
  (i) saturation pressure of gold defined by current temperature of the gold SS
    and later down to (ii) critical pressure of water 218 bars.
 The last instant is marked by the black circle in Figure 3B.
 After that instant the bubble in water begins to form.
 Equation of state for water was taken from the database \cite{database:H2O};
   in two-temperature hydrodynamics (2T-HD) simulations we use the Hugoniot adiabatic curve of water \cite{database:H2O};
     thus in these 2T-HD simulations we cannot follow the two-phase decay of water.


 Contact pressure drop is shown in Figure 3B as the dependence ``CB'' (contact boundary).
 The dependence ``CB'' has three stages.
 At the first one the pressure decrease is slowed down by arrival of momentum and mass
   from foam arriving (accreting) to the atmospheric layer.
 The first stage lasts at the subnanosecond time scale.


 Later on the foam is exhausted, thus the foam momentum support of atmosphere pushing water is ceased,
  and the rate of decrease of contact pressure becomes higher, see the curve ``CB'' in Figure 3B.
 This is the second stage - the fast unsupported deceleration of the atmospheric layer by inertia of water.
 It covers the time interval between subnanoseconds and few nanoseconds in Figure 3B.





 The third stage follows after the second stage.
 It begins at 3-4 nanoseconds after launch of expansion, see the dependence ``CB'' in Figure 3B.
 At this stage the pressure drop at the gold-water contact again is slowed down.
 Now the reason is the dynamical support by saturated pressure of gold
   evaporated into the back space - the space behind the atmospheric layer,
     i.e. into the spatial gap between the layer and the bottom of the crater.
 At the third stage the contact pressure is defined by temperature of the atmosphere.
 Temperature of the atmosphere is shown by the curve ``T'' in Figure 3B.
 The curve ``T'' relates to the right vertical axis in Figure 3B.
 The saturated pressure of gold corresponding to this temperature
   is shown by dependence ``saturated vapor'' in Figure 3B.



 Integrating the increment of RTI (see reviews \cite{Inogamov:RT-engl:1999,Inogamov:RT-rus:1999})
   along the calculated trajectory of the contact boundary,
     we find the resulting gain of an amplitude of contact perturbation shown in Figure 3C.
 Increasing of amplitude of linear perturbation with wavelength $\lambda$ along the contact
  continues up to the moment when surface tension cutting of increment at wavelength $\lambda_\sigma (t)$
   achieves the considered wavelength $\lambda.$
 In Figure 3C we see that significant amplification of perturbations takes place
   during the time interval from the launch to the instant of cutting the increment.
 The non-linear growth of the RTI is considered in the next Chapter.




  \section{5. MD-MC SIMULATION OF GOLD/WATER INTERACTION}

 Layered structure before and after formation of foam are shown in Figures 4 and 5.
 Results of MD-MC simulation are shown in Figure 6.
 Large scale simulation is based on our own interatomic potentials for gold and water \cite{Zhakhovskii:2009,project:EAM}.
 Simulation confirms details found in 2T-HD runs:
  wide cavitation zone, formation of atmosphere, accretion of foam onto atmosphere, RT instability of contact.
 MD-MC gives pictures of non-linear growth of RTI, see also \cite{LZ:2017a,LZ:2017b}.
 It shows enlargement of the leading lateral scale typical for RTI \cite{Inogamov:RT-engl:1999,Inogamov:RT-rus:1999}
  and shows decay of jets into droplets.
 Decay of deceleration, values of surface tension and viscosity play important roles
   for development of linear and non-linear RTI.

 The droplets before separation are linked to the long jets.
 Process of decay into droplets in liquid (Plateau-Rayleigh instability) and solid states \cite{FLAMN-QE:2017}
  begins with existence of elongated droplets and droplets with the long tail.
 If freezing is fast than this tail remains at the solidified droplets.
 Such kind of mysterious droplets were seen in experiments - see Figures 2 and 4 in \cite{Shafeev:2009-NP-tail}.

  \section{CONCLUSION}

 We consider different aspects of foaming of molten metal under action of intense ultrashort laser action.
 Dynamics of expansion of cavitation zone in case of large fluences few times above ablation threshold
  is analyzed in Sections 2-4.
 This foam plays definitive role in gold-water dynamic interaction.
 1D structure of gold-water flow is studied in Section 4. 
 It is found that atmosphere like layer from gold forms at the contact with water.

 The atmosphere from one side is decelerated by water, while from another side it is attacked
   by fragments of foam falling to atmosphere.
 This mass and momentum flux decreases deceleration of atmosphere.
 Linear stability analysis based on 1D trajectory of contact from 1D 2T-HD simulations
   points out to the limited growth of RT perturbations (Figure 3C).
 Direct MD-MC simulation allows to understand non-linear details of development of instability (Chapter 5).


\vspace{-2pt}
\section{ACKNOWLEDGMENTS}

 This research was support by Russian Science Foundation (RSCF) project  No. 14-19-01599.



\end{document}